\documentclass{svjour3}                     

\smartqed  
\def\makeheadbox{\relax}
\usepackage[top=2.54cm, bottom=2.54cm, left=1.91cm, right=1.91cm]{geometry}
\usepackage{latexsym}
\usepackage{graphicx}
\usepackage{amssymb}
\usepackage{amsmath}
\usepackage{subfigure}
\usepackage{multirow}
\usepackage{rotating}
\usepackage[table]{xcolor}
\usepackage[T1]{fontenc}
\usepackage[polutonikogreek,english]{babel}
\usepackage{babel}
\usepackage{epstopdf}
\usepackage{tikz}
\usepackage{dsfont}
\usepackage{textcomp}
\usepackage{mathrsfs}
\usepackage{comment}
\usepackage{microtype}
\usepackage{epsfig}
\usepackage{cite}
\usepackage{blindtext}
\usepackage{lmodern}
\usepackage{listings}
\usepackage{placeins}
\usepackage{amsmath,amssymb,amsfonts}

\usepackage{algorithmicx}
\usepackage{algorithm}
\usepackage{algpseudocode}
\usepackage{pifont}
 \usepackage[flushleft]{threeparttable}

\DeclareGraphicsExtensions{.pdf,.png,.jpg}
\journalname{Multimedia tools and applications}

\begin{document}
\lstset{language=c++,breaklines=true,showspaces=false,showtabs=false}
\title{A Secure and Improved Multi Server Authentication Protocol Using Fuzzy Commitment}

\author{Hafeez Ur Rehman  \and Anwar Ghani \and Shehzad Ashraf Chaudhry  \and Mohammed H. Alsharif \and Narjes Nabipour}
\authorrunning{Rehman  et al.}

\institute{Hafeez  Ur Rehman  \at
             Department of Computer Science and Software Engineering, International Islamic University Islamabad,Pakistan \\
             Tel.: +92-312-901823\\
             \email{hafeezkami@gmail.com}           
\and
           	  Anwar Ghani* \at
             Department of Computer Science and Software Engineering, International Islamic University Islamabad,Pakistan \\
             \email{anwar.ghani@iiu.edu.pk}
\and
    				 Shehazad Ashraf Chudhry \at
             Department of Computer Engineering, Faculty of Engineering and Architecture Istanbul Gelisim University Istanbul, Turkey \\
             \email{ashraf.shehzad.ch@gmail.com}
\and
    				 Mohammed H. Alsharif \at
             Department of Computer Engineering, Faculty of Engineering and Architecture Istanbul Gelisim University Istanbul, Turkey \\
             \email{moh859@gmail.com} 
\and
    				 Narjes Nabipour* \at
             Institute of Research and Development, Duy Tan University, Da Nang, Vietnam \\
             \email{narjesnabipour@duytan.edu.vn} 
						}

\date{\today}
\maketitle
\begin{abstract}

Very recently, Barman et al. proposed a multi-server authentication protocol using fuzzy commitment. The authors claimed that their protocol provides anonymity while resisting all known attacks. In this paper, we analyze that Barman et al.’s protocol is still vulnerable to anonymity violation attack and impersonation based on stolen smart attack; moreover, it has scalability issues. We then propose an improved and enhanced protocol to overcome the security weaknesses of Barman et al.'s scheme. The security of the proposed protocol is verified using BAN logic and widely accepted automated AVISPA tool. The BAN logic and automated AVISPA along with the informal analysis ensures the robustness of the scheme against all known attacks. 
\end{abstract}
\section{Introduction}
\label{sec:intro}

The multi-server environment provides convenient and suitable online services as unlike conventional single server authentication (SSA), the multi-server environment provides single sign-on without registering with multiple servers and keeping the multiple secrets of passwords and identities. The multi-server works using the centralized trusted registration authority, responsible for registering the servers and users, in return it enables both the servers and users to get hassle free communication with each other. The users keeps only a secret password and an identity. The common use of a multi-server environment requires an efficient and robust user authentication protocol to establish a secure connection between both the requesting user and service providers.

In 1981, Lamport \cite{1lamport1981} presented the first authentication protocol based on a server database containing the passwords of each registered user. Due to storage of the verifier in server database Lamport's protocol is subjected to the stolen verifier attack. Over time, many researchers proposed their protocols to resolve the issues of stolen verifier attack \cite{2fan2005robust,3juang2008robust}.Wu et al.’s \cite{4wu2012secure} presented a smart card-based authentication protocol; later He et al. \cite{5debiao2012more} noticed that the protocol of Wu is vulnerable to insider attack and impersonation attack. Wu et al.’s \cite{4wu2012secure} then presented an improved and enhanced protocol based on He et al.'s protocol. later Zhu et al.\cite{6zhu2012efficient} found that the protocol of He et al. still has some weaknesses like offline password guessing attack.\par

Anticipating the failure and/or unsuitability of two factor authentication protocols, many researchers proposed fingerprint-based three factor authentication protocols to enhance the security\cite{7lee2002fingerprint,8fan2009provably,9li2010efficient}. Lee et al. \cite{7lee2002fingerprint} presented fingerprint-based authentication. Lee et al. enhanced the security using three factors including: 1)smart card, 2)fingerprint minutiae, and 3)user password. Later  Lin et al.'s \cite{10lin2004flexible} claimed that Lee et al.'s protocol has weaknesses against spoofing and masquerade attacks. So they proposed an enhanced protocol based on Lee et al.'s protocol. Regretfully, Mitchell et al. \cite{11mitchell2005security} noticed that Lin et al.'s protocol still has some weaknesses. Amin et al.'s \cite{12amin2015novel} presented a novel protocol for multi-server architecture in which the authors claimed that their protocol provides security against the known attacks. Later Das et al.'s \cite{13das2015secure} noticed that Amin et al.'s protocol suffers from many attacks. Das et al. also proved that Amin et al.'s protocol does not reinforce the biometric update. \par
Mir and Nikooghadam\cite{14mir2015secure} presented an enhanced biometrics-based authentication protocol and claimed their protocol provides security against well-known attacks like (user anonymity and untraceability, impersonation attacks, Online password guessing attacks, etc.) Later Chaudhry et al. \cite{15chaudhry2018enhanced} noticed that  Mir and Nikooghadam \cite{14mir2015secure} suffers from user anonymity attack as well as stolen smart attack. Unfortunately, Qi et al. \cite{16qi2018new} claimed  Chaudhry et al.'s \cite{15chaudhry2018enhanced} protocol still has some weaknesses including non-resilience against denial of service attack; moreover, protocol in \cite{15chaudhry2018enhanced} is lacking perfect forward secrecy. \par

In 2016, Wang et al. \cite{17wang2016cryptanalysis} proposed another biometric-based multi-server authentication and key agreement protocol based on Mishra et al.’s protocol. Wang et al. claimed their protocol provides various security features along-with user revocation/re-registration and biometric information protection. Soon, AG Reddy et al. \cite{18reddy2017design} showed that Wang et al.'s \cite{17wang2016cryptanalysis} protocol is vulnerable to server impersonation, user impersonation and insider attacks, as their protocol share user credential to the server. Qi et al.'s \cite{19qi2017efficient} proposed yet another key-exchange authentication protocol and claimed it to provide security against well-known attacks. later AG Reddy et al.'s \cite{20reddy2018privacy} noticed some vulnerabilities like session key leakage attack, user impersonation attack, insider attack, and user anonymity in the protocol of Qi et al. \par

Very recently, Barman et al.\cite{21barman2018provably} proposed a provably secure multi-server authentication protocol using fuzzy commitment. The authors in \cite{21barman2018provably} claimed that their protocol provides various security features like confidentiality of user identity/biometric data, mutual authentication and session key establishment between user and servers, besides this authors also claimed their protocol to provide security against the known attacks. However, the in-depth analysis in this article shows that the protocol of Barman et al. is facing some serious security threats. It is to show that the protocol proposed by Barman et al. is vulnerable to anonymity violation attack and impersonation attack based on stolen smart-card. Moreover, their protocol is not practicable owing to the scalability issues. Then we propose an improved and enhanced protocol to overcome the security weaknesses of Barman et al.'s protocol. We analyze the security of our proposed protocol through formal and informal analysis. In the formal analysis, we used a BAN Login and widely accepted AVISPA tool, a well known and widely accepted automated tool for security analysis . The informal security features analysis also shows the robustness of the proposed protocol.

\section{Preliminaries}
\label{sec:preli}
A brief review of the basics relating to fuzzy commitment technique, one-way hash function, error correction coding, and revocable template generation, is solicited in this section.

\subsection{Fuzzy Commitment}
\label{sec:fuz}
The fuzzy commitment as proposed by Juels and Wattenberg \cite{26juels1999fuzzy} is a method to hide the secrets under the witness and then release the conceal secrets later in the presence of a witness. In the Registration/enrollment phase a randomly generated key $K_{c}$ is cipher with codeword $C_{w} = \aleph_{enc}(K_{c})$. $\aleph_{enc}$ is an error correction technique and it helps in a noisy channel to recover equivalent match. When a user imprints his biometric then the binary string is generated against that biometric $C_{T_u}$ used to conceal the key with binary string through XOR operation [$C_{T_u} \oplus C_{w} = H_{public}$]. The system contain only $H_{public}$ and the hash of key $(h(K_{c}))$. In the authentication phase this  $H_{public} $ is available, so every legitimate user imprints his/her biometric to unlock $C_{w}$.
\subsection{Hash Function}
Hash function h: X $\xrightarrow{}$ Y is deterministic mapping set  X = $\{0,1\}^*$ of strings having variable length to another set Y = $\{0,1\}^t$ of strings of fixed length, properties include:
\begin{itemize}
	\item The input value say, $a \in X $ it is easy to compute h(a), in polynomial times moreover, h(.) function is deterministic in nature.
	\item The change in input value $a \in X$ results in a completely uncorrelated with h(a).
	\item  \textbf{$One-Way$ $property:$} It is difficult to find the actual message $x$ given the message digest $h(a)$ of $a \in X$.
	\item  \textbf{$Weak-Collision$ $resistant$ $property$:} Any given value input $a \in X$. it is difficult to find another $a^{*} \in X$ such that $h(a) = h(a^{*})$.
	\item \textbf{$Strong-Collision$ $resistance$ $property$:} $h(a) = h(a^{*})$ for any $a$, $a^{*} \in X$ and $a \neq a^{*}$, this property states that, it is also difficult to find any two inputs $a, a^{*} \in X$ such that $a \neq a^{*}$ with $h(a) = h(a^{*})$. 
\end{itemize}
\subsection{Revocable Template Generation}
\label{sec:temp_gen}

A revocable template \cite{27ratha2007generating}, provides the privacy and Revocability of user biometric. By using transformation parameter $TP_{u}$ and transformation function, $f(\cdot)$, user biometric data  is convert into a cancel able template $CT_{u} = f(BIO_{u}, TP_{u})$, properties includes:
\begin{enumerate}
	\item \textbf{Collision-free property:}  If $CT_{u} = f(BIO_{u}, TP_{u})$ and $CT_{k} = f(BIO_{k}, TP_{k})$, then $CT_{u} \neq CT_{k}$. for $BIO_{u} \neq BIO_{k}$.  Moreover, if $CT_{n} = f(BIO, TP_{n})$ and $CT_{m} = f(BIO, TP_{m})$,  then $CT_{n} \neq CT_{m}$ for $TP_{n} \ne TP_{m} $.
	\item \textbf{Intra-user variability property :} This property states; two different templates $CT_{u} = f(BIO_{u}, TP_{u})$, $CT^{'}_{u} = f(BIO^{'}_{u}, TP_{u})$ can be generated form same fingerprint. 
	\item \textbf{Revocation of biometric:} If user biometric is comprised, then new template can be generated by using new transformation parameter $TP^{new}_{u}$ with same transformation function $f(\cdot)$. 
	\item \textbf{User confidentiality:} Cancel-able template should protect the confidentiality of user, moreover template should protect the information about original biometric of a user. 
\end{enumerate}
\subsection{Error Correction Technique}
\label{sec:corr_tec}
In the biometric template, the intra-user variation is considered an error. To remove the errors in the user biometric template, error correction technique \cite{28hao2006combining} is used for noisy biometric image. In the time of enrollment/Registration $CT_{enrol_{u}} = f(BIO_{enrol_{u}},TP_{u})$ is generated, which is match with query template  $CT_{query_{u}} = f(BIO_{query_{u}},TP_{u})$, at the authentication time. So the difference can be calculated through Hamming distance $ e = HamDis(CT_{enrol_{u}}, CT_{query_{u}})$.
\subsection{Adversarial Model}
\label{sec:adv}

According to the well known and widely accepted Dolev-Yao threat (DY) model \cite{22dolev1983security}, an attacker not only listen to the communication between two participants but also the attacker can change the entire message or delete the message as well on open channel. An attacker can also extract the secret credential of legitimate user form stolen smart card through power analysis attack\cite{23kocher1999differential, 24messerges2002examining}. Second adversarial model is  Canetti and Krawczyk model (CK-model). In authentication and key exchange protocol, it is considered as De-facto standard. According to the (CK-adversary model)\cite{25canetti2001analysis}, it is not only fallowed  Dolev-Yao threat (DY) model but the adversary is also able to get the session key and session states as well.

\subsection{Our contributions}
\label{sec:contri}

\begin{enumerate}
	\item We have cryptanalyzed the recent fuzzy commitment based multi-server authentication protocol proposed by Barman et al.'s \cite{21barman2018provably} to find out its security issues and vulnerabilities.
	\item We proposed an improved and enhanced authentication protocol based on Barman et al.'s \cite{21barman2018provably} 
	\item The security of the proposed protocol is checked through BAN logic and widely accepted AVISPA. 
	\item The security discussion and security features comparisons of the proposed protocol with related protocols including Barman et al.'s protocol is explained.
	\item We have also provided the comparative computation and communication costs analysis of the proposed protocol with competing related protocols
\end{enumerate}

\subsection{Notations}
The notations used in this paper are provided in fig 1.
\FloatBarrier

\FloatBarrier
\begin{figure}[ht]
	\centering
	\scalebox{0.9}{
		\begin{tabular}{|l l|}
			\hline
			Symbols & Representations \\
			\hline	\hline
			$U_{u}$,$S_{k}$ & user and server \\ 
			$SID_{k}$  & identity of server\\ 
			$ID_{u},PW_{u},BIO_{u}$ & identity, password and biometric of $U_{u}$ \\ 
			$CT_{u},TP_{u},f(.)$ & cancel-able template, transformation parameter \\ & and transformation function of $U_{u}$ \\  
			$RC$ & trusted registration center \\   
			$X_{c}$ & secret/private key of $RC$ \\ 
			${XR_{k}}$ & shared keys between $S_{k}$ and $RC$ \\
			$E_{X_{c}},D_{X_{c}}$ & encryption and decryption using private key of $RC$\\ 
			$R_{cu}$ & user's random number\\  
			$H_{u}$ & fuzzy commitment helper data \\  
			$SK_{u,_k}$ & session key between user $U_{u}$, $S_{k}$ \\  
			$PSK_{k}$ & secret/private key of $S_{k}$\\ 
			$h(.)$ &  hash function \\  
			$R_{u},r_{n},R_{s}$ & random number generated by $U_{u}$, $RC$, $S_{k}$ \\   
			$T_{1},T_{2},T_{3},$ & time stamped generated by $U_{u},RC, S_{k}$ \\   
			$T_{u}$ & time bound generated by $S_{k}$ \\  
			$\Delta$T & time delay \\   
			$\oplus,\parallel$ &  $(XOR)$ and string concatenation operator \\  
			$\aleph_{enc}(.), \aleph_{dec}(.)$ & encoding and decoding operator ,\\ & of the error correction technique \\  
			$SC_{u}$, $A_{adv}$ & smart card and adversary \\
			\hline
		\end{tabular}
	} 
	\scriptsize
	\label{table:NGuide}
	\caption{Notations}
\end{figure}
\section{Review Of Barman et al's Protocol }
\label{sec: pre_scheme}
This section briefly reviews Barman et al.'s protocol \cite{21barman2018provably}. The six phases of the protocol are detailed in following subsection:
\subsection{Server Registration Procedure}
\label{sec:ser_reg}
In Barman et al protocol, all servers $S_{k}: (1 \leq k \leq n)$, where $n$ denotes the total number of servers in the network. Initially, all servers \textit{$S_{k}$}, $(1 \leq k \leq n)$ will registered with $RC$. Every server \textit{$S_{k}$} selects their particular identity $SID_{k}$ and dispatches a registration request to the $RC$.
 $RC$ sends a secret key $PSK_{K}$ = $h(SID_{k}||X_{c})$ to each $S_{k} (1 \leq k \leq n)$. $RC$ may also consider another $n_{'}$ servers, which will register themselves with the $RC$ near in future. Therefore, the $RC$ chooses their identities $SID_{S}$ and generates the shared keys  $PSK_{S}$ = $h(SID_{S} || X_{c})$ for $n+1 \leq S\leq n + n^{'}$  The server identities (for $n + n^{'}$ server) along with their corresponding key pairs ${(SID_{k},PSK_{k})|1 \leq k \leq n + n_{'}}$ are stored in $RC$ database.

\subsection{User Registration Procedure}
\label{sec:user_reg}
The detail steps of the user registration phase are defined below:
\begin{enumerate}
	\item Initially, every user $U_{u}$ needs to register with the $RC$ to gets the services, via a protected channel. $U_{u}$ select  a unique user identity $ID_{u}$, and password $PW_{u}$, a transformation parameter $T_{P_{u}}$ and a random number $Rc_{u}$. $U_{u}$ also imprint $BIO_{u}$.
	
	\item $U_{u}$ produce the cancel-able biometric template using transformation functions $CT_{u}$ = $f(BIO_{u}, TP_{u})$ and computes $RPW_{u}$ = $h(PW_{u} || CT_{u})$, $r_{u}$ = $h(Rc_{u} ||ID_{u} || PW_{u})$. $U_{u}$. $U_{u}$ then generates a random secret $k_{u}$ and sends the registration request $ \langle ID_{u}, RPW_{u} \oplus	k_{u}\rangle $ to the $RC$, via a protected channel. 
	
	\item After checking validity of $ID_{u}$. $RC$ computes $US_{k}$ = $h(ID_{u} ||PSK_{k})$, $AM_{k}$ = $US_{k} \oplus (RPW_{u}  \oplus k_{u})$, $SV_{k}$ = $h(SID_{k} || PSK_{k})$ and $BM_{k}$ = $SV_{k} \oplus $\(RPW_{u} \oplus k_{u}\) for 1 $\leq k \leq (n + n^{'})$. $RC$ issues a smart card $SC_{u}$ having $\{(SID_{k}, AM_{k}, BM_{k})$$| 1 \leq k \leq (n + n^{'})\}$ and sends it to $U_{u}$, via a protected channel.
	
	\item Using error correction technique $\varepsilon$. $U_{u}$ encodes $Rc_{u}$ produced codeword $R_{cod} = \varepsilon_{enc}(Rc_{u})$, computes $H_{u}$ = $CT_{u} \oplus R_{cod}$, $R$ = $h(Rc_{u})$ and $P$ = $h(r_{u})$. $U_{u}$ then computes $AM_{uk}$ = $(AM_{k} \oplus k_{u}) \oplus r_{u}$ and $BM_{uk}$ = $(BM_{k} \oplus k_{u}) \oplus r_{u}$ for $ 1 \leq k \leq (n + n^{'} )$.\par
	$U_{u}$ stores $\{(AM_{uk}, BM_{uk})\} | 1 \leq k \leq (n + n^{'}),$ $TP{u}, H_{u}, R, $ $P, h(\cdot), \aleph_{enc}(\cdot), \aleph_{dec}(\cdot)\}$ in smart card $SC_{u}$. $U_{u}$ cancels the $Rc{u}, BIO_{u}, CT{u}, r_{u}, AM_{k} $ and $BM_{k}$ for security reasons.
	
\end{enumerate}


\subsection{Login Procedure}
\label{sec:log_pro}
The detail steps of login request are:
\begin{enumerate}
	\item $U_{u}$ inserts the smart card into the terminal and provides the credentials $ID_{u}, PW_{u}$ and $BIO^{'}_{u}$ for authentication.
	
	\item  The smart card $SC_{u}$ generates the cancel-able fingerprint $CT^{'}_{u} = f(BIO^{'}_{u}, TP_{u})$, and extracts $R^{'}_{cod} = H_{u} \oplus CT^{'}_{u} $ and then decodes $R^{'}_{cod}$ using error correction technique, $Rc^{'}_{u}  = \aleph_{dec}(R^{'}_{cod})$. $SC_{u}$ compares both values, $h(Rc^{'}_{u})$ with $R$ which is stored in $SC_{u}$. If they are equal than proceed further else terminate the session.
	
	\item $SC_{u}$ computes $r^{'}_{u} = h(Rc_{u} || ID_{u} || PW_{u})$ and checks if $h(r^{'}_{u}) = h(r_{u})$, proceed further otherwise terminate the session.
	
	\item $SC_{u}$ computes $US_{k} = AM_{uk} \oplus h(PW_{u} || CT_{u}) \oplus r^{'}_{u} = h(ID_{u} || PSK_{k}) $ and $SV_{k} = BM_{uk} \oplus h(PW_{u} || CT_{u}) \oplus r^{'}_{u} = h(SID_{k} || PSK_{k})$. $SC_{u}$ selects a random number $R_{u}$, generates current time stamp $T_{1}$, and computes $M^{'}_{1} = h(ID_{u}||US_{k}),M^{'}_{2} = ID_{u} \oplus h(SV_{k}||T_{1}), M_{3} = M_{1} \oplus R_{u}, M_{4} = h(ID_{u}||M^{'}_{1}||M^{'}_{2} || T_{1}||R_{u})$.
	
	\item Finally, $SC_{u}$ sends the request $ \langle M^{'}_{2}, M^{'}_{3}, M^{'}_{4}, T_{1} \rangle $ to the server $S_{k}$. 
\end{enumerate}
\subsection{Mutual Authentication and Key Agreement Procedure}
\label{sec:auth_keyagre}
The mutual authentication and key agreement consist of following steps:
\begin{enumerate}
	\item $S_{k}$ receives login request $ \langle M^{'}_{2}, M^{'}_{3}, M^{'}_{4}, T_{1} \rangle $ at time $T^{'}_{1}$ and after computing the time delay, $|T^{'}_{1} - T_{1}|$. Computes $M^{'}_{5} = M^{'}_{2} \oplus h(h(SID_{k}||$ $PSK_{k})||T_{1}), M^{'}_{6} = h(M^{'}_{5}||h(M^{'}_{5}||PSK_{k})) $ $M^{'}_{7} = M^{'}_{3} \oplus M^{'}_{6} = R_{u}$ and $M^{'}_{8} = h(M^{'}_{5}||M^{'}_{6}||M^{'}_{2}||T_{1}||M^{'}_{7})$. Check if $M^{'}_{8} \neq M^{'}_{4}$, $S_{k}$ cancel the login request, else proceed further.
	\par
	\item $S_{k}$ select a random number  $R_{s}$ and generates $T_{3}$ then computes $M^{'}_{9} = h(h(M^{'}_{5}||PS_{k})||R_{u}) \oplus R_{s}$, and session key $SK_{uk} = $$h(M^{'}_{5}||h(SID_{k}||PSK_{k})||R_{u}||R_{s}||T_{1}||T_{3})$ and $M^{'}_{10} = h(h(M^{'}_{5}||PSK_{k})||SK_{uk}||T_{3}||R_{s})$ sends $ \langle M^{'}_{9},M^{'}_{10},T_{3} \rangle $ to $U_{u}$.
	
	\item  The $U_{u}$ receives $ \langle M^{'}_{9}, M^{'}_{10},T_{3} \rangle $. After checking the delay $|T^{}_{3} \leq T_{c}|$. $SC_{u}$  computes $R^{'}_{s} = M^{'}_{9} \oplus h(US_{k}||R_{u})$, the session key $SK^{'}_{uk} = h(ID_{u}||SV_{k}$$||R_{u}||R_{s}||T_{1}||T_{3})$ shared with $S_{k}$ and $M^{'}_{11} = h(US_{k}||SK^{'}_{uk}||T_{3}||R^{'}_{s})$. $SC_{u}$ check the condition if $M^{'}_{11} \neq M^{'}_{10}$ terminated. Otherwise, the session key $SK_{uk}$ is established between $U_{u}$ and $S_{k}$.
\end{enumerate}
\subsection{Password and Biometric Template Update Procedure}
\label{sec:pass_bio_up}

$U_{u}$ provides the current credentials $ID_{u},PW_{u}$ $BIO_{u}$ and extracts feature $BIO^{'}_{u}$ from the $BIO_{u}$. $SC_{u}$ then computes $CT^{'}_{u} = f(BIO^{'}_{u},TP_{u})$ and $Rc^{'}_{u} = \aleph_{dec}(H_{u} \oplus CT^{'}_{u})$ and then check if $h(Rc^{'}_{u}) = R, SC_{u}$ further computes $r^{'}_{u} = h(Rc^{'}_{u}||ID_{u}||PW_{u})$ check if $h(r^{'}_{u}) = P$ proceed further otherwise terminate. $SC_{u}$ then request to the user $U_{u}$ to modify their password and biometric template.

\begin{enumerate}
	\item To update the password, $U_{u}$ inputs $PW^{new}_{u}$. Then, $SC_{u}$ computes $r^{new}_{u} = h(Rc^{'}_{u}||ID_{u}|| $ $ PW^{new}_{u}), AM^{new}_{uk}$ = $AM_{uk} \oplus r^{'}_{u} \oplus r^{new}_{u} = h(ID_{u}||PSK_{u}) \oplus h(PW^{'}_{new}||CT_{u}) \oplus$  $h(Rc^{'}_{u} $ $ ||ID_{u}||PW^{new}_{u})$, $ BM^{new}_{uk}$ = $BM_{uk} \oplus r^{'}_{u} \oplus r^{new}_{u} = $ $  h(SID_{k}||PSK_{k}) \oplus h(PW^{new}||CT_{u}) \oplus $ $ h(Rc^{'}_{u} $ $ ||ID_{u} $ $ ||PW^{new}_{u})$ for $1 \leq k \leq (n + n^{'})$ and $P^{new} = h(r^{new}_{u}). SC_{u}$ updates its parameters $\{AM_{uk},BM_{uk},\}$ with the newly computed values $\{AM^{new}_{uk} $,$ BM^{new}_{uk} $ $, P^{new}\}$ and stored in the $SC_{u}$.
	
	\item To update the biometric template, $SC_{u}$ request $U_{u}$ for a new transformation parameter $TP_{u}$. $SC_{u}$ have the old $TP_{u}$ and then set new  $TP^{new}_{u} = TP_{u}$ and new cancel-able template is produce  $CT^{new}_{u} = f(BIO^{'}_{u},TP^{new}_{u})$. $SC_{u}$ also computes $RPW^{new}_{u} = h(PW_{u}||CT^{new}_{u})$, $AM^{new}_{uk} = AM_{uk} \oplus RPW_{u} \oplus RPW^{new}_{u}$ = $h(ID_{u}|| $ $ PSK_{k}) \oplus h (PW_{u} $ $ ||CT^{new}_{u} ) r^{'}_{u}, BM^{new}_{uk} = BM_{uk} \oplus RPW_{u} \oplus$ $RPW^{new}_{u}$ = $h(SID_{k}|| $ $ PSK_{k}) \oplus h(PW_{u}||CT^{new}_{u})\oplus r^{'}_{u}$, and the new helper data $H^{new}_{u} = CT^{new}_{u} \oplus \aleph_{enc}(Rc^{'}_{u})$. Accordingly, the information $\{AM_{uk}, BM_{uk}, H_{u}\}$ is replaced by $\{AM^{new}_{ij}$  $BM^{new}_{uk}, H^{new}_{u}\}$ stored in the $SC_{u}$ .
\end{enumerate} 
\subsection{Smart Card Revocation Procedure}
\label{sec:card_rev}

If the $SC_{u}$ of a authorized $U_{u}$ is damaged, lost or stolen, then $U_{u}$ can get a new $SC_{u}$ from the $RC$. $U_{u}$ provides $ID_{u}$ and $PW_{u}$ and to imprint $BIO_{u}$, Steps are: 	
\begin{enumerate}
	\item $U_{u}$ computes $CT^{'}_{u} = f(BIO_{u},TP_{u})$ and $RPW_{u} =h(PW_{u}||CT^{'}_{u})$,  $U_{u}$  generates a random number $k^{'}_{u}$, then computes a parameter $RPW^{'}_{u} = RPW_{u} \oplus k^{'}_{u}$ and then sends the request $ \langle ID_{u}, RPW^{'}_{u} \rangle $ to the $RC$ via a protected channel for a new $SC^{new}_{u}$ 
	
	\item $RC$ computes $AM_{k} = h(ID_{u}||PSK_{k}) \oplus RPW^{'}_{u}, BM_{k} = h(SID_{k}||PSK_{k}) \oplus RPW^{'}_{u}$ for $k = 1,2,,,,,,(n+n^{'})$ and issue a new $SC^{new}_{u}$ containing $\{(SID_{k}, AM_{k}, BM_{k})| 1 \leq k \leq n +n_{'} \}$.  $SC^{new}_{u}$ sends to these parameter to $U_{u}$ via a protected channel.
	
	\item $U_{u}$ generates a new random number $R^{new}_{u}$ and computes $r_{u} = h(R^{new}_{u} || ID_{u} || PW_{u}), H^{new}_{u} = CT^{'}_{u} \oplus \aleph_{enc} (R^{new}_{u}), AM_{uk} = (AM_{k} \oplus k^{'}_{u}) \oplus r_{u}, BM_{uk} = (BM_{k} \oplus k^{'}_{u}) \oplus r_{u}, R = h(Rc^{new}_{u}), P = h(r_{u})$ and stores these values in $SC^{new}_{u},$ memory. $U_{u}$ also stores $\{TP_{u}, \aleph_{enc}(\cdot), \aleph_{dec}(\cdot), h(\cdot)\}$ in $SC^{new}_{u} $ memory.
\end{enumerate}
\section{Cryptanalysis of Barman et al. protocol}
\label{sec:Crypt}

Barman et al.'s protocol \cite{21barman2018provably} provides multi-server based authentication protocol using a fuzzy commitment approach. The in depth analysis proves that the  protocol entails serious security flaws as described in following subsections:
\subsection{Incomplete Login Request}
\label{sec: loginreq}
The login message, $\{M^{'}_{2} , M^{'}_{3}, M^{'}_{4}, T_{1}\}$ sent by user $U_{u}$ to the server $S_{k}$ is incomplete, because the identity of server $SID_{k}$ is not included in the login request, which is the most important parameter while communication\cite{29lwamo2019suaa} and without the server identity, the $RC$ cannot direct the request of $U_u$ to his intended server. This crucial mistake may be treated as typing mistake. The protocol can only work if the login message contains the identity of the server.
\subsection{User Anonymity Violations Attack}
\label{sec: violatt}

Here, we show that the protocol of Barman et al. is vulnerable to user anonymity violation attack. Let $U_{a}$ be a legal but dishonest user of the system and wants to violate user anonymity. In the Mutual Authentication phase of Barman et al.'s protocol user $U_{u}$ sends the message $\{M^{'}_{2}, M^{'}_{3}, M^{'}_{4}, T_{1}, SID_{k}\}$  to the server $SID_{k}$ on public channel. During the communication, let $U_{a}$ intercepts the message and from $M^{'}_{2} = ID_{u} \oplus h(SV_{k}\| T_{1})$, $U_{a}$ can easily extract the $ID_{u}$ of every users. Because all the users connected to the $SID_{k}$ has $SV_{k}$(secret identifier generated by RC for $SID_{k}$) which are stored in the smart card. $U_{a}$ can extract the identity of user as follows:

\newcounter{UAcounter}
\begin{list}{Step AV \arabic{UAcounter}:~}{\usecounter{UAcounter}}
\item $U_{u}$ sends the login message to $SID_{k}$. During the communication, let user $U_{a}$ intercepts the message $\{M^{'}_{2} , M^{'}_{3}, M^{'}_{4}, T_{1}, SID_{k}\}$.

\item $U_{a}$ using his own smart card, enters his credentials including: $ID_a$, $PW_a$ and $BIO_a$. $U_a$ extracts $\{BM_{ak},AM_{ak}\}$ pair from his own smart card and then computes $CT_a=f(BIO_a,TP_a)$, $R^{'}_{cod}=H_a\oplus CT_a$, $Rc^{'}_a  = \aleph_{dec}(R^{'}_{cod})$, $r_a=h(Rc_u||ID_a||PW_a)$, similar to login steps. $U_a$ then computes: 

\begin{align}
&US_{k_a}=AM_{ak}\oplus h(PW_a||CT_a)\oplus r_a\\
&SV_{k} = BM_{ak} \oplus h(PW_{a} || CT_{a}) \oplus r^{'}_{a} = h(SID_{k} || PSK_{k})\\
&Z=h(SV_{k}||T_1)
\end{align}

\item Based on $SV_{k}$, $Z$ and the $M^{'}_{2}$ from login request, $U_a$ computes:

\begin{align}
ID_u&=M^{'}_{2}\oplus Z \label{UID}
\end{align}

\end{list}

In Eq.\ref{UID}, the $ID_u$ is the real identity of $U_u$. Therefore, $U_a$ has successfully broke the user anonymity.


\subsection{User Impersonation Attack based on stolen smart card} 
\label{sec:stolen}
Using the stolen smart card of some user say $U_a$, another legal but dishonest user of the system can launch user impersonation attack in Barman et al.'s protocol. Let $U_a$ be a legal user, gets his card $SC_{a}$ containing $\{SID_{k}, AM_{a_k},BM_{a_k} | 1 \leq k \leq (n + n^{'})\}$ along with $\{TP_{a}, H_{a}, P, h(\cdot), \aleph_{enc}, \aleph_{dec}\}$ and steals the smart card $SC_u$ of $U_a$ performs following steps to impersonate on behalf of $U_u$:
\newcounter{gscounter}
\begin{list}{Step ISC \arabic{gscounter}:~}{\usecounter{gscounter}}
\item $U_{a}$ enter his credential  $ID_{a}, PW_{a}$ and biometric $BIO_{a}$. $U_{a}$ computes $US_{k}, CT^{'}_{a}, r^{'}_{a} $, $SV_{k} = BM_{uk} \oplus h(PW_{a} || CT_{a}) \oplus r^{'}_{a} = h(SID_{k} || PSK_{k})$. As $SV_{k}$ is common in all smart cards.\\

\item Extracts $AM_{k_{u}}  = US_{k_{u}} \oplus (RPW_{u} \oplus k_{u}) $ and $BM_{uk} = SV_{k} \oplus (RPW_{u} \oplus k_{u})$ form $U_{u}$'s  stolen smart card $SC_u$. \\

\item $U_{a}$ using $SV_{k}$ computes:
\begin{align}
		&X= AM_{k_{u}} \oplus BM_{k_{u}} = \{US_{k_{u}} \oplus (RPW_{u} \oplus k_{u})\} \oplus \{SV_{k} \oplus (RPW_{u} \oplus k_{u})\}\\
		&X = US_{k_{u}} \oplus SV_{k}\\
		&US_{k_{u}} = X \oplus SV_{k}
	\end{align}
\item $U_{a}$ has $SV_{k}$ and $US_{k_{u}}$ of $U_{u}$ with $ID_{u}$. $U_{u}$ generates a random number $R_{u}$ and time stamp $T_{1}$ computes:

	\begin{align} 
		M^{'}_{1} &= h(ID_{u} || US_{k})\\
		M^{'}_{2} &= ID_{u} \oplus h(SV_{k} || T_{1})\\
		M^{'}_{3} &= M^{'}_{1} \oplus R_{u}\\
		M^{'}_{4} &= h(ID_{u} || M^{'}_{1} || M^{'}_{2} || T_{1} || R_{u})
	\end{align}

\item $U_{a}$ sends the login request message $ \langle M^{'}_{2}, M^{'}_{3}$, $M^{'}_{4}, T_{1}$, $SID_{k} \rangle $ to the $S_{k}$. $S_{k}$ receives the login request $ \langle M^{'}_{2}, M^{'}_{3}, M^{'}_{4}, T_{1}, SID_{k} \rangle $ after checking time delay, $|T^{'}_{1} - TS_{1}|$, computes following:

\begin{align}
		&M^{'}_{5} = M^{'}_{2} \oplus h(h(SID_{k}||PSK_{k})||T_{1}) = (ID_{u})\\
		&M^{'}_{6} = h(M^{'}_{5}||h(M^{'}_{5}||PSK_{k}))\\
		&M^{'}_{7} = M^{'}_{3} \oplus M^{'}_{6} = R_{u}\\
	  &M^{'}_{8} = h(M^{'}_{5}||M^{'}_{6}||M^{'}_{2}||T_{1}||M^{'}_{7})
	\end{align}

\item $S_k$ checks if $M^{'}_{8} = M^{'}_{4}$, $U_u$ will pass this test because $M^{'}_{8}$ and $M^{'}_{4}$ both have same values. Therefore user $U_{a}$ pass test on behalf of $U_{u}$. $S_{k}$ selects a nonce $R_{s}$, generates current time stamp $T_{3}$, and computes:
\begin{align} 
&M^{'}_{9} = h(h(M^{'}_{5}||PS_{k})||R_{u}) \oplus R_{s}\\
&SK_{uk} =h(M^{'}_{5}||h(SID_{k}||PSK_{k})||R_{u}||R_{s}||T_{1}||T_{3})\label{sSKey}\\
& M^{'}_{10} = h(h(M^{'}_{5}||PSK_{k})||SK_{uk}||T_{3}||R_{s})
\end{align}

\item Then, $S_{k}$ sends $ \langle M^{'}_{9},$ $ M^{'}_{10}, T_{3} \rangle $ to $U_{a} $. $U_{a}$ receives the authentication request message $ \langle M^{'}_{9}, M^{'}_{10}, T_{3} \rangle $ at time $T^{'}_{3}$. $U_{a}$ computes following:

\begin{align}
		&R_{s} = M^{'}_{9} \oplus h(US_{k}||R_{u})\\
		&SK^{'}_{uk} = h(ID_{u}||SV_{k}||R_{u}||R_{s}||T_{1}||T_{3})\label{SKey}\\
		&M^{'}_{11} = h(US_{k}||SK^{'}_{uk}||T_{3}||R_{s})
\end{align}

\end{list}
The session key as computed by $U_a$ in Eq. \ref{SKey} is same as computed by $S_k$ in Eq.\ref{sSKey}. Therefore, $U_a$ has succesffuly established a secure connection with $S_k$ by impersonating on behalf of $U_a$.

\subsection{Scalability problem}
\label{sec:scalbility}

In the registration phase of Barman et al.'s protocol smart card stores $AM_{k}$. As in multi-server environment, there may be several servers and users. So it is inefficient to store $(AM_{k})$ against every server within smart card due to its small magnetic chip which has limited storage. This protocol is not practical, suppose we have n servers, so we need to store $US_{k}$ and $SV_{k}$ of n servers within the smart card, each of size 160 bits. For large number of servers like 100, the bits stored for $US_{k}$ and $SV_{k}$ in the smart card are 32000 bits, which can be problematic due to its storage restrictions. Moreover, authors did not mention the procedure to update the smart card if some new servers are added, $AM_{uk}$ = $(AM_{k} \oplus k_{u}) \oplus r_{u}$ and $BM_{uk}$ = $(BM_{k} \oplus k_{u}) \oplus r_{u}$ for $ 1 \leq k \leq (n + n^{'} )$.
\section{Proposed Protocol}
\label{sec:pro_sch}
This section details the proposed scheme consisting of three entities including, users, servers and the Registration Center (RC). The details are in following subsections:
\subsection{Server Registration Phase}
\label{sec:proserReg}

Every \textit{S$_{k}$} along with its particular identity $SID_{k}$ must send a registration request to the $RC$, if they are willing to provide services to the legitimate users $U_{u}$. RC computes $X_{RS_k} = h(SID_{k}||X{c})$ and $M_{k} = E_{X_c}(X_{RS_{k}}) $ and stores $(SID_{k},E_{X_c}(X_{RS_k}))$ in the database of $Rc$ and send the share key to the server $(X_{RS_k})$.
\subsection{User Registration Phase }
$U_u$ chooses $ID_{u}, PW_{u}, TP_{u}$, then imprint $BIO_{u}$ and selects random number $N_{1}$. $U_u$ computes $CT_{u} = f(BIO_{u}, TP_{u}), A_{u} = h(N_{1}||PW_{u}||ID_{u}||CT_{u})$ and sends $A_{u}, ID_{u}$ to the $RC$. On receiving $RC$ computes $X_{u} = h(ID_{u}||X_{c})$ and $Y_{u} = Xu \oplus A_{u}$ then generate a random number $r_{o}$ and compute the pseudo identity $PID_{u} = E_{X_c}(ID_{u}||r_{o})\oplus A_{u}$, then store $Y_{u}, PID_{u}, h(.)$ in smart card. $RC$ sends the smart card to user using some secure channel. On receiving smart card, $U_{u}$ computes $R_{c} = \aleph_{enc}(Rc_{u}), H_{u} = CT_{u}\oplus R_{cod}, R = h(Rc_{u})$, $r_{u} = (Rc_{u}||ID_{u}||PW_{u})$, $P = h(r_{u})$ and $E_{u} = N_{1} \oplus r_{u}$. $U_u$ stores $ \{ TP_{u}, H_{u}, R, P, h(.)$, $\aleph_{enc}(\cdot), \aleph_{dec}(\cdot), Y_{u}, $ $ PID_{u}, E_{u} \}$ in the smart card.
\FloatBarrier
\begin{figure}[ht]
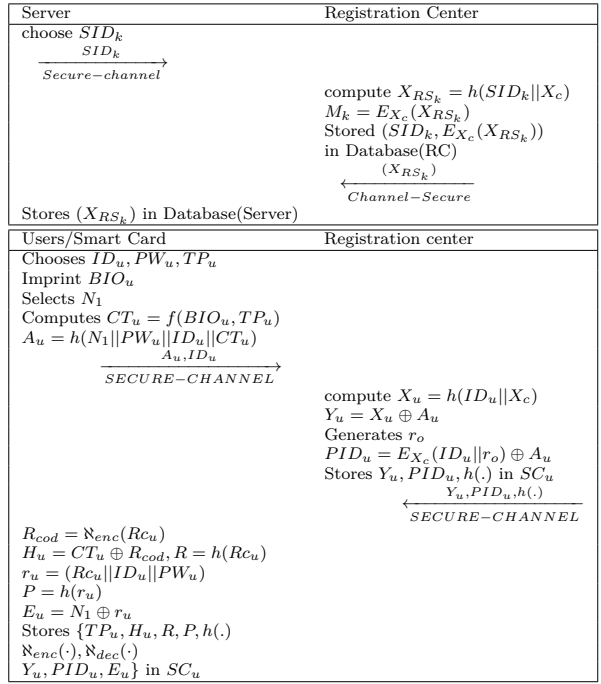

	\centering
	\scalebox{0.8}{
		\begin{tabular}{|l l|}
			\hline
			Server & Registration Center \\
			\hline
			choose $SID_{k}$ & \\
			$\phantom{h}\xrightarrow[Secure-channel]{SID_{k}}$ & \\
			& compute $X_{RS_k} = h(SID_{k}||X_{c})$ \\
			& $M_{k} = E_{X_{c}}(X_{RS_k})$  \\
			& Stored $(SID_{k}, E_{X_{c}}(X_{RS_k}))$ \\
			& in Database(RC)  \\
			& $\phantom{h}\xleftarrow[Channel-Secure]{(X_{RS_k})}$ \\
			Stores $(X_{RS_k})$ in Database(Server) & \\
			\hline
			\hline
			Users/Smart Card  & Registration center \\
			\hline
			Chooses $ID_{u},PW_{u},TP_{u}$ & \\
			Imprint $BIO_{u}$ & \\
			Selects $N_{1}$ & \\
			Computes $CT_{u} = f(BIO_{u},TP_{u})$ & \\
			$A_{u} = h(N_{1}||PW_{u}||ID_{u}||CT_{u}) $ & \\
			$\phantom{hhhhhhh}\xrightarrow[SECURE-CHANNEL]{A_{u},ID_{u}}$& \\
			& compute $X_{u} = h(ID_{u}||X_{c}) $ \\
			& $Y_{u} = X_{u} \oplus A_{u}$ \\
			& Generates $r_{o}$ \\
			& $PID_{u} = E_{X_{c}}(ID_{u}||r_{o}) \oplus A_{u} $ \\
			& Stores $Y_{u},PID_{u},h(.)$ in $SC_{u}$ \\
			& $\phantom{hhhhhhh}\xleftarrow[SECURE-CHANNEL]{Y_{u},PID_{u},h(.)}$ \\
			$R_{cod} = \aleph_{enc}(Rc_{u})$ & \\
			$H_{u} = CT_{u}\oplus R_{cod}, R = h(Rc_{u})$ & \\
			$r_{u} = (Rc_{u}||ID_{u}||PW_{u})$ & \\
			$P = h(r_{u})$ & \\
			$E_{u} = N_{1} \oplus r_{u} $ & \\
			Stores $ \{ TP_{u},H_{u},R,P,h(.)$ & \\ 
			$\aleph_{enc}(\cdot),\aleph_{dec}(\cdot) $& \\ 
			$ Y_{u},PID_{u},E_{u} \}$ in $SC_{u}$ & \\
			\hline
		\end{tabular}
	}
	\caption{Registration phase of Sever and User}
\end{figure}
\FloatBarrier
\subsection{Login and Authentication Phase}
\label{sec:LoginAuth}

The following steps define the login and authentication phase briefly:
\newcounter{0bscounter}
\begin{list}{Step AP \arabic{0bscounter}:~}{\usecounter{0bscounter}}
	
	\item  User need to insert the smart card provides the credentials $ID_{u}, PW_{u}, BIO^{'}_{u}$ and calculate $CT^{'}_{u} = f(BIO^{'}_{u}, TP_{u})$, $R^{'}_{cod} = H_{u} \oplus CT^{'}_{u}$, $Rc^{'}_{u} = \aleph_{dec}(R^{'}_{cod})$, and check if $h(Rc^{'}_{u}) \neq R$, terminate the session, otherwise calculate $r^{'}_{u} = h(Rc^{'}_{u}||ID_{u}||PW_{u})$, and check again if $h(r^{'}_{u}) \neq h(r_{u})$ terminate the session, else compute $N_{1} = (E_{u} \oplus r_{u})$, $A^{'}_{u} = h(ID_{u}||PW_{u}||N_{1}||CT_{u})$, $X_{u} = (Y_{u} \oplus A^{'}_{u})$, $DID_{u} = (PID_{u}\oplus A^{'}_{u})$,  generate a random no $R_{u}$ and time stamp $T_{1}$, and to get the services of server needs the address $SID_{k}$, and computes $G_{u} = R_{u} \oplus h(X_{u}||ID_{u}||SID_{k}||T_{1})$, $H_{u} = h(ID_{u}||G_{u}||X_{u}||R_{u}||T_{1}||SID_{k})$, sends $\{DID_{u}, H_{u}$, $ G_{u}, T_{1}, SID_{k}\}$ to the Registration on public channel.

	\item $RC$ receives the login request and checks the time delay $(T_{c} - T_{1} \leq \delta T)$. $RC$ decrypts $(ID_{u}||r_{o}) = D_{X_c}(PID_{u})$ using $X_{c} $ and computes $X_{u} = h(ID_{u}||X_{c})$ $R_{u} = G_{u}\oplus h(X_{u}||ID_{u}||SID_{k}||T_{1})$ $H^{'}_{u} = h(ID_{u}||G_{u}||X_{u}||R_{u}||T_{1}||SID_{k})$. $RC$ then check $ H^{'}_{u} \stackrel{?}{=} H_{u} $ if not true, terminates the session. Otherwise, $RC$ verify user successfully, and then $RC$ extracts $X_{RS_{k}} $ from verifier table of $RC$, and generate time stamp $T_{2}$  computes $X^{'}_{u} = h(X_{u}||ID_{u}||SID_{k}||T_{1})$, $H_{R_c} = $ $ h(X_{RS_{k}}||X^{'}_{u}||ID_{u}||SID_{k}||T_{2})$, and encrypt the parameters $ (X^{'}_{u}, R_{u}, ID_{u}, H_{R_c}, SID_{k}, T_{1}, T_{2})$ using share secret key $X_{RS_{k}}$ and sends $E_{X_{RS_{k}}} $, $ (X^{'}_{u}$ $R_{u}, ID_{u}, H_{R_c}, SID_{k}, T_{1}, T_{2})$ ,$SID_{k}$ to the server over public channel.

\item On receiving the message, $S_k$ after checking the time delay $(T_{c} - T_{2}\leq \delta T )$, decrypts $ D_{X_{RS_{k}}}(X^{'}_{u}, R_{u}, ID_{u}, H_{R_c},$ $  SID_{k}, T_{1})$ using the shared key ${X_{RS_{k}}}$. $S_k$ then computes $H^{'}_{R_{c}} = h (X_{RS_{k}}||$ $ X^{'}_{u}||ID_{u}||SID_{k}||T_{2})$ and checks the equality $H^{'}_{R_{c}} \stackrel{?}{=} H_{R_{c}} $ if condition is true, $S_k$ verifies $RC$ successfully. Further $S_k$ generates $R_{s}$, $T_{3}$ and computes $M_{x} = R_{s}\oplus h (ID_{u}||X^{'}_{u}||R_{u}||T_{3})$ $H^{''}_{R_{c}} = h(R_{s}||M_{x}|| T_{u}||ID_{u}|| $ $ T_{3})$. $S_k$ further sends $\{M_{x}, H^{''}_{R_{c}}, T_{3}, T_{u},\}$  to the $RC$, which in turn checks $(T_{c} - T_{3} \leq \delta T )$ and in successful verification computes $R_{s} = M_{x}\oplus (ID_{u}||X^{'}_{u}||R_{u}||T_{3})$ $H^{'''}_{R_{c}} = h(R_{s}||M_{x}||T_{u}||ID_{u}||T_{3})$. $RC$ then checks  $H^{'''}_{R_{c}} \stackrel{?}{=} H^{''}_{R_{c}} $ terminates the session on success; otherwise, computes new dynamic identity $RID_u=E_{X_c}(ID_u||r_n)\oplus R_{s}$ for $U_u$ and  forwards $\{M_{x}, H^{''}_{R_{c}}, T_{3}, T_{u}, RID_u\}$ to the legitimate user $U_u$.
		
\item $U_u$ on receiving the message, checks $T_{3} \leq \delta T_{c} $ and on success computes $R_{s} = M_{x}\oplus (ID_{u}||X^{'}_{u}||R_{u}||T_{3}) $, $ H^{''''}_{R_{c}} =  h(R_{s}|| M_{x}||T_{u}||ID_{u}||T_{3})$ and checks whether $H^{''''}_{R_{c}} $ $ \stackrel{?}{=} H^{''}_{R_{c}}$ if true then session key $SK_{uk} = $ $h(X^{'}_{u}$$||ID_{u}||SID_{k}||R_{s}||R_{u})$ is established between user and server.
\end{list}

\subsection{Password and Biometric Update Process}
In this section, we also proposed the password change and biometric template update process of our protocol, the $U_{u}$ will need to log in successfully to change their current password and update their biometric template, The detailed steps are described below:
\newcounter{1bscounter}
\begin{list}{Step CPB \arabic{1bscounter}:~}{\usecounter{1bscounter}}
	\item $U_{u}$ provides the credentials $ID_{u}, PW_{u}$, and $BIO_{u}$ after inserting the smart-card into a card reader to login. $BIO^{'}_{u}$ is extracted from the captured $BIO_{u}$. $SC_{u}$ then computes $CT^{'}_{u} = f(BIO^{'}_{u}, TP_{u})$ and 
	$R^{'}_{cu} = \varepsilon_{dec}(H_{u}\oplus CT^{'}_{u})$. Checks if $h(R^{'}_{cu}) = R$, then $SC_{u}$ computes $r^{'}_{i} = h(R^{'}_{cu}||ID_{u}||PW_{u})$, and check if $h(r^{'}_{i}) = P$, smart card then asks users $U_{u}$ to change their password and update their biometric template.

\FloatBarrier
\begin{figure}[htb]
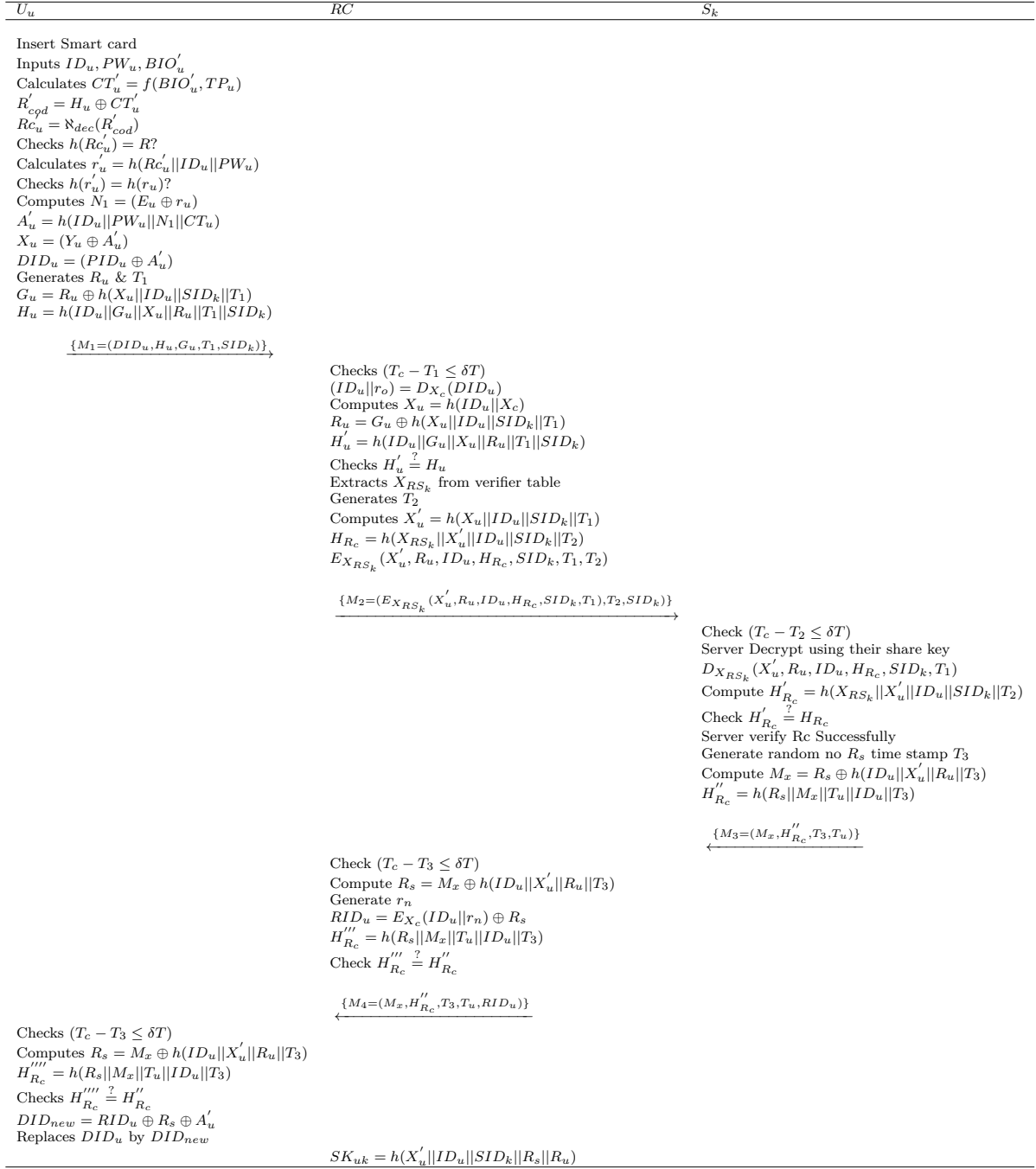

	\centering
	\scalebox{0.8}{
		\begin{tabular}{l l l}
			\hline 
			$U_{u}$  & $RC$ & $S_{k}$ \\
			\hline
			&&\\
			Insert Smart card & & \\
			Inputs $ID_{u},PW_{u},BIO^{'}_{u}$ & & \\
			Calculates $CT^{'}_{u} = f(BIO^{'}_{u},TP_{u})$ & & \\
			$R^{'}_{cod} = H_{u} \oplus CT^{'}_{u} $ &&\\
			$Rc^{'}_{u} = \aleph_{dec}(R^{'}_{cod})$ &&\\
			Checks $h(Rc^{'}_{u}) = R?$ &&\\
			Calculates $r^{'}_{u} = h(Rc^{'}_{u}||ID_{u}||PW_{u})$ &&\\
			Checks $h(r^{'}_{u}) = h(r_{u})? $ &&\\
			Computes $N_{1} = (E_{u} \oplus r_{u})$ && \\
			$A^{'}_{u} = h(ID_{u}||PW_{u}||N_{1}||CT_{u})$ && \\
			$X_{u} = (Y_{u} \oplus A^{'}_{u})$ && \\
			$DID_{u} = (PID_{u}\oplus A^{'}_{u})$ && \\
			Generates $R_{u}$ \& $T_{1}$ && \\
			$G_{u} = R_{u} \oplus h(X_{u}||ID_{u}||SID_{k}||T_{1})$ && \\
			$H_{u} = h(ID_{u}||G_{u}||X_{u}||R_{u}||T_{1}||SID_{k})$ && \\
			&&\\
			$\phantom{hhhhh}\xrightarrow[]{\{M_{1}=(DID_{u},H_{u},G_{u},T_{1},SID_{k})\}}$ && \\
			& Checks $(T_{c} - T_{1} \leq \delta T)$ & \\
			& $(ID_{u}||r_{o})= D_{X_{c}}(DID_{u})$ & \\
			& Computes $X_{u} = h(ID_{u}||X_{c})$ & \\
			& $R_{u} = G_{u}\oplus h(X_{u}||ID_{u}||SID_{k}||T_{1})$ & \\
			&$H^{'}_{u} = h(ID_{u}||G_{u}||X_{u}||R_{u}||T_{1}||SID_{k})$ & \\
			& Checks  $ H^{'}_{u} \stackrel{?}{=} H_{u} $ & \\
			& Extracts $X_{RS_{k}} $ from verifier table & \\
			& Generates $T_{2}$ & \\
			& Computes $X^{'}_{u} = h(X_{u}||ID_{u}||SID_{k}||T_{1})$ & \\
			& $H_{R_c} = h(X_{RS_{k}}||X^{'}_{u}||ID_{u}||SID_{k}||T_{2})$  & \\
			& $E_{X_{RS_{k}}}(X^{'}_{u},R_{u},ID_{u},H_{R_c},SID_{k},T_{1},T_{2})$ & \\
			&&\\
			&$\phantom{}\xrightarrow[]{\{M_{2}=(E_{X_{RS_{k}}}(X^{'}_{u},R_{u},ID_{u},H_{R_c},SID_{k},T_{1}),T_{2},SID_{k})\}}$ & \\
			&& Check $(T_{c} - T_{2} \leq \delta T)$\\
			&& Server Decrypt using their share key \\
			&& $D_{X_{RS_{k}}}(X^{'}_{u},R_{u},ID_{u},H_{R_c},SID_{k},T_{1})$ \\
			&& Compute $H^{'}_{R_{c}} =  h(X_{RS_{k}}||X^{'}_{u}||ID_{u}||SID_{k}||T_{2}) $ \\
			&& Check $H^{'}_{R_{c}} \stackrel{?}{=} H_{R_{c}} $ \\
			&& Server verify Rc Successfully \\
			&& Generate random no $R_{s}$ time stamp $T_{3}$  \\
			&& Compute $M_{x} = R_{s}\oplus h(ID_{u}||X^{'}_{u}||R_{u}||T_{3}) $ \\
			&& $H^{''}_{R_{c}} = h(R_{s}||M_{x}||T_{u}||ID_{u}||T_{3}) $ \\
			&&\\
			&& $\phantom{}\xleftarrow[]{\{M_{3}=(M_{x}, H^{''}_{R_{c}},T_{3},T_{u})\}}$  \\
			& Check $(T_{c} - T_{3} \leq \delta T )$  & \\
			& Compute $R_{s} = M_{x}\oplus h(ID_{u}||X^{'}_{u}||R_{u}||T_{3})$ & \\
			& Generate $r_n$&\\
			& $RID_u=E_{X_c}(ID_u||r_n)\oplus R_{s}$ & \\
			& $H^{'''}_{R_{c}} = h(R_{s}||M_{x}||T_{u}||ID_{u}||T_{3})$ & \\
			& Check $H^{'''}_{R_{c}} \stackrel{?}{=} H^{''}_{R_{c}} $ & \\
			&&\\
			& $\phantom{}\xleftarrow[]{\{M_{4}=(M_{x}, H^{''}_{R_{c}},T_{3},T_{u}, RID_u)\}}$ & \\
			Checks $(T_{c} - T_{3} \leq \delta T)$ &&  \\
			Computes $R_{s} = M_{x}\oplus h(ID_{u}||X^{'}_{u}||R_{u}||T_{3})$  && \\
			$H^{''''}_{R_{c}} = h(R_{s}||M_{x}||T_{u}||ID_{u}||T_{3})$ &&  \\
			Checks $H^{''''}_{R_{c}} \stackrel{?}{=} H^{''}_{R_{c}}$ && \\
			$DID_{new}=RID_u\oplus R_s \oplus A^{'}_{u}$\\
			Replaces $DID_u$ by $DID_{new}$&&\\
			& $SK_{uk} = h(X^{'}_{u}||ID_{u}||SID_{k}||R_{s}||R_{u})$ & \\
			
			\hline 
		\end{tabular}
	}
	\caption{Login and Authentication phase}
\end{figure}
\FloatBarrier

	\item For password change, $SC_{u}$ asks $U_{u}$ for a new password. $U_{u}$ inputs the new password $PW^{new}_{u}$. $SC_{u}$ computes $r^{new}_{u} = h(R^{'}_{cu}||ID_{u}||PW^{new}_{u}), E^{new}_{u} = N_{1}\oplus r^{new}_{u}$  and $P^{new} = h(r^{new}_{i})$. $SC_{u}$ updates its parameters stored $\{TP_{u}, H_{u}, R,$ $ P^{new}, h(\cdot), \varepsilon_{enc}(\cdot), \varepsilon_{dec}(\cdot), Y_{u}, PID_{u}, E^{new}_{u}\}$ in smart card in its memory.
	
	\item To update the biometric template, $SC_{u}$ asks $U_{u}$ for a new transformation parameter $TP^{new}_{i}$. The new cancel-able template is generated as $CT^{new}_{i} = f(BIO_{u}, TP^{new}_{i})$, and the new helper data $H^{new}_{i} = CT^{new}_{i}\oplus \varepsilon_{enc}(R^{'}_{ci})$ and are stored in $SC_{u}$.
\end{list}

\subsection{Smart Card Revocation Procedure}
In this section, we proposed the smart card revocation, if the $SC_{u}$ of the legitimate user $U_u$ is damaged, lost or stolen, then $RC$ will issue the new smart card. For this process, the user provides their credential $ID_{u}, PW_{u}, BIO_{u}$. The following steps are essential to complete this procedure:

\newcounter{2bscounter}
\begin{list}{Step SCR \arabic{2bscounter}:~}{\usecounter{2bscounter}}
	\item  $U_u$ computes $CT^{'}_{i} = f(BIO_{i}, TP_{i})$ and generates a $160$-bit secret $N^{'}_{1}$, then computes $A^{'}_{u} = h(N^{'}_{1}||PW_{u}||ID_{u}||$ $ CT^{'}_{u})$, and transmits the request message $\{A^{'}_{u},ID_{u}\}$ to the $RC$ via a protected channel for $SC^{new}_{u}$.
	\item   $RC$ computes $X_{u} = h(ID_{u}||Xc), Y^{'}_{u} = X_{u}\oplus A^{'}_{u}$, generate random $r^{'}_{o}$ and computes $PID^{'}_{u} = E_{X_{c}}(IDu||r^{'}o)\oplus A^{'}_{u}$ store $Y^{'}_{u}, PID^{'}_{u}, h(.)$ in $SC_{u}$, then issue a $SC^{new}_{i}$ containing the credentials $,Y_{u}, PID^{'}_{u}, h(.)$. $SC^{new}_{i}$ is then sent to $U_{i}$ via a protected channel. 
	\item $U_u$  computes $r^{'}_{u} = h(Rc^{new}_{i}||ID_{u}||PW_{u})$, $H^{u}_{new} = CT^{'}_{u} \oplus \varepsilon_{enc}(Rc^{new}_{u}),  , R = h(Rc^{new}_{u}), P = h(r_{u})$ and stores these values in $SC^{new}_{i}$ memory. 
\end{list}
\section{Security Analysis}
\label{sec:sec_ana}

\begin{figure}[ht]
	\centering
	\includegraphics[width=0.5\linewidth]{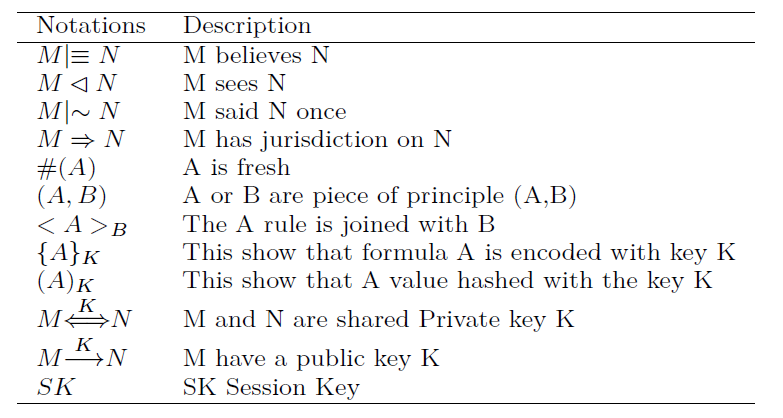}
	\caption{Notations}
	\label{fig:bannotations}
\end{figure}

In this section, we analyze our protocol using widely accepted Burrows-Abadi-Needham (BAN) logic \cite{30burrows1989logic}, used to check the mutual authentication between the user $U_{u}$, server $S_{k}$ and registration center $RC$,
 The notation used in the BAN logic is given in fig \ref{fig:bannotations}. 
\FloatBarrier
\FloatBarrier
\subsection{Rules of BAN-Logic}
\label{sec:RulesBAn}

The rules of authentication protocol are clearly mentioned in \cite{30burrows1989logic}, which illustrate that if only one rule is violate, then the entire protocol is consider as flawed. Rules are define table in \ref{sec:RulesBANLOG}:\newline

\FloatBarrier
\begin{table}[ht]
	\centering
	\begin{tabular}{ c c c }
		\hline
		& Rules & Definition \\ 
		\hline
		Rule 1 & Message Meaning & $\frac{M|\equiv M\overset{K}\longleftrightarrow N.M\lhd <A>_K} {M|\equiv N|\sim A}$ \\ 
		Rule 2& Nonce Verification & $\frac{M |\equiv\#(A),M |\equiv N|\sim A}{M |\equiv N|\equiv A}$ \\  
		Rule 3& Jurisdiction & $\frac{M |\equiv N\Rightarrow A, M|\equiv N|\equiv A}{M|\equiv A}$ \\
		Rule 4& Acceptance Conjuncatenation & $\frac{M |\equiv\ A,M |\equiv B}{M |\equiv (A,B)}$ \\
		Rule 5 & Freshness Conjuncatenation & $\frac{M |\equiv\#(A)} {M |\equiv\#(A,B)}$ \\
		Rule 6 & Session Key & $\frac{M |\equiv\#(A),M|\equiv N\equiv A} {M|\equiv\ M\overset{K}\longleftrightarrow N}$ \\
		\hline   
	\end{tabular}
\caption{Rules of BAN-Logic}
\label{sec:RulesBANLOG}
\end{table}
\FloatBarrier

\subsection{Assumptions} 

\begin{itemize}
	\item A1:$U_{u}|\equiv \#(R_{u},T_{1})$ 
	\item A2:$S_{k}|\equiv \#(R_{s},T_{3},T_{u})$ 
	\item A3:$RC|\equiv \#(T_{2})$ 
	\item A4:$RC|\equiv U_{u}|\equiv \#(R_{u},T_{1})$
	\item A5:$RC|\equiv S_{k}|\equiv \#(R_{u},T_{1})$
	\item A6:$RC|\equiv RC\overset{X_{RS_{k}}}\longleftrightarrow S_{k} $
	\item A7:$S_{k}|\equiv RC\overset{X_{RS_{k}}}\longleftrightarrow S_{k} $
	\item A8:$RC|\equiv U_{u} \Rightarrow R_{u} $  
	\item A9:$RC|\equiv S_{k} \Rightarrow R_{s},T_{u} $ 
	\item A10:$U_{u}|\equiv RC|\equiv S_{k}|\equiv \#(R_{s},T_{3},T_{u})$ 
	\item A11:$S_{k}|\equiv RC|\equiv U_{u}|\equiv \#(R_{u},T_{1})$ 
	\item A12:$RC|\equiv U_{u}|\equiv S_{k}|\overset{SK_{uk}}\longleftrightarrow U_{u}$ 
	\item A13:$RC|\equiv S_{k}|\equiv U_{u}|\overset{SK_{uk}}\longleftrightarrow S_{k}$ 
\end{itemize}
\subsection{Goals}
\begin{itemize}
	\item $U_{u}|\equiv U_{u}\overset{SK_{uk}}\longleftrightarrow S_{k}$ 
	\item $S_{k}|\equiv U_{u}\overset{SK_{uk}}\longleftrightarrow S_{k}$
\end{itemize}
\textbf{The Idealized Form of Messages} Four message are used in session key agreement process which are:
\begin{itemize}
	\item \textbf{Messages(1)}$U_{u} \xrightarrow{} RC$:$\{DID_{u}, H_{u}, G_{u}, T_{1}, SID_{k}\}$
	\item \textbf{Messages(2)}$RC \xrightarrow{} S_{k}$:$\{E_{X_{RS_{k}}}(X^{'}_{u}, R_{u}, ID_{u}, H_{R_c}, SID_{k},$ $ T_{1}, T_{2}),  SID_{k}\}$
	\item \textbf{Messages(3)}$S_{k} \xleftarrow{} RC$:$\{M_{x}, H^{''}_{R_{c}}, T_{3}, T_{u}\}$
	\item \textbf{Messages(4)}$RC \xleftarrow{} U_{u}$:$\{M_{x}, H^{''}_{R_{c}}, T_{3}, T_{u}, RID_{u}\}$
\end{itemize}
$M_{1}$:- ${U_{u} \lhd\{DID_{u}, H_{u}, G_{u}, T_{1}, SID_{k}}\}$ or $U_{u} \lhd \{ID_{u},r_{o}\}_{X_{c}}$, $ h(ID_{u}, G_{u}, X_{u}, R_{u}, $ $ T_{1}, SID_{k})$,  $ R_{u} \oplus $ $ h(X_{u}, ID_{u}, $ $ SID_{k}, T_{1}),\\ T_{1}, SID_{k}> $..........Eq.(A)\\
$M_{2}$:- $S_{k} \lhd< \{X^{'}_{u}, R_{u}, ID_{u}, H_{R_c}, SID_{k}, T_{1}, T_{2}\}_{X_{c}}, SID_{k}>$ or
$S_{k} \lhd $ $\{h(X_{u}, ID_{u}, T_{1})$, $ h(X_{RS_{k}}, X^{'}_{u}, $ $ ID_{u}, SID_{k}, T_{2})$, $R_{u}, ID_{u}, SID_{k}, T_{1}, T_{2}\}_{X_{c}}, SID_{k},>$......Eq.(B)\\
$M_{3}$:- $RC \lhd $ $<M_{x}, H^{''}_{R_{c}}, T_{3}, T_{u}>$ or $R_{c} \lhd $ $<R_{s}\oplus h(ID_{u}, X^{'}_{u}, R_{u}, T_{3})$, $h(R_{s}, M_{x}, T_{u}, ID_{u}, T_{3})$, $ T_{3}, T_{u}> $.....Eq.(C)\\
$M_{4}$:- $U_{u} \lhd $ $<M_{x}, H^{''}_{R_{c}}, T_{3}, T_{u}, RID_{u}>$ or $R_{c} \lhd $ $<U_{i}\oplus h(ID_{u}, X^{'}_{u}, R_{u}, T_{3})$, $h(R_{s}, M_{x}, T_{u}, ID_{u}, T_{3})$, $ T_{3}, T_{u}, \{{ID_{u},r_{n}\}_{X_{c}}} $.....Eq.(D)\\
\subsection{Protocol Analysis}

The main security proofs are consist of the following steps:
\begin{itemize} 
	\item From message "$M_{1}$" using \textbf{(A1, A12)} and \textbf{Rule-1}
	we get\\
	BN1: $RC|\equiv U_{u}\sim R_{u} $ \\
	\item Using \textbf{A1} and \textbf{Rule-2} on "BN1" we get\\
	BN2: $RC|\equiv U_{u}|\equiv R_{u}$\\ 
	\item Using \textbf{A8} and \textbf{Rule-3} on "BN2" we get.\\
	BN3: $RC|\equiv R_{u}$\\$RC$ believes that $R_{u}$ is fresh based on \textbf{A3} and \textbf{Rule-5} \\
	\item From message "$M_{2}$" using \textbf{A7} and \textbf{Rule-1} we get\\
	BN4: $S_{k}|\equiv RC \sim X_{RS_{k}}$\\
	\item Using (\textbf{A7, A11}) and \textbf{Rule-3} on "BN4" we get\\
	BN5: $S_{k}\equiv X_{RS_{k}}$\\
	Server believes that $X_{RS_{k}}$ is secret parameter which is only known to $S_{k}$ and $RC$. Using (\textbf{A7, A11}) and \textbf{Rule-6} on "BN5" we get\\
	BN6: $S_{k}|\equiv U_{u} \overset{SK_{uk}}\longleftrightarrow S_{k}$ Goal-1 achieved\\
	\item From message "$M_{3}$" using  (\textbf{A2, A5}) and \textbf{Rule-1} we get\\
	BN6: $RC|\equiv S_{k} \sim (R_{s}, T_{3}, T_{u})$\\
	\item Using \textbf{A2} and (\textbf{Rule-2}) on "BN6" we get\\
	BN7: $RC|\equiv S_{k}|\equiv (R_{s}, T_{3}, T_{u})$ \\
	\item Using  \textbf{A9} and \textbf{Rule-3} on "BN7" we get\\
	BN8: $RC|\equiv (R_{s}, T_{3}, T_{u})$\\
	\item From message "$M_{4}$" using (\textbf{A4, A10}) and \textbf{Rule-1} we get\\
	BN9: $U_{u}|\equiv RC\sim (R_{s}, T_{3}, T_{u}, RID_{u}) $ \\
	\item Using (\textbf{A10}) and \textbf{Rule-3} on "BN9" we get\\
	BN10: $U_{u}|\equiv RC|\equiv (R_{s}, T_{3}, T_{u}, RID_{u})$\\ 
	\item Using \textbf{A9} and \textbf{Rule-3} on "BN10" we get\\
	BN11: $U_{u}|\equiv (R_{s}, T_{3}, T_{u}, RID_{u})$\\
	$U_{u}$ believes that  $(R_{s}, T_{3}, T_{u}, RID_{u})$ are fresh. $R_{s}$ is an important parameter for session key agreement process.
	\item Using \textbf{A1} and \textbf{Rule-6} on "BN11" we get\\
	BN12: $U_{u}\equiv U_{u}\overset{SK_{uk}}\longleftrightarrow S_{k}$. Goal-2 achieved\\
\end{itemize}
\section{Discussion on Functional Security}
\label{sec: diss_funsec}
Following subsection solicits brief discussions on several security features and resistance to known attacks provided by the proposed scheme.

\subsection{Anonymity and Untraceability}

In the authentication protocol, user anonymity and untraceability are substantial aspects and if anonymity is broken, an adversary $A_{adv}$ can easily recover sensitive information of the legitimate user like his current location, moving tracks, a personal record and social circle, etc. In the registration phase $RC$ encrypt the identity with random number $E_{X_{c}}(ID_{u}||r_{o})$ by using his own secret key $X_{c}$. $SC_{u}$ does not store this pseudo identity directly, as it is hidden by $PID_{u}$, So even if the smart card were stolen by $A_{adv}$ he will still be incapable to get the identity of the user. Moreover, after each successful authentication request, this pseudo-identity is dynamically changed. Therefore, the proposed protocol provides anonymity and untreceability.
\subsection{Impersonation Attacks}

To act as $RC$ an $A_{adv}$ required the secret key $X_{c}$ of $RC$, which is hash with user identity $h(ID_{u}||X_{c})$, to compute the session key $SK = h(X^{'}_{u}||ID_{u}||SID_{k}||R_{s}||R_{u})$ an $A_{adv}$ also requires to first compute $X_{u} =  h(ID_{u}||X_{c})$. In addition $X_{u}$ is also used in the construction of $RC$ signature that is, $X^{'}_{u} = h(X_{u}||ID_{u}||SID_{k}||T_{1})$. So without secret key $X_{c}$ an $A_{adv}$ does not impersonate themselves as $RC$. Similarly to act as legitimate user an $A_{adv}$ will required a valid login request that is,$ \{DID_{u},H_{u},G_{u},T_{1},SID_{k}\}$. To get all these values an $A_{adv}$ needs the user credential like password $PW_{u}$ as well as biometric $BIO_{u}$.
\subsection{Replay Attack}

Our protocol combat replay attack against all the login and authentication messages. Suppose an $A_{adv}$ replays a past message that is $\{ DID_{u}, H_{u}, G_{u}, T_{1}, SID_{k}\} $. then on receiving side $RC$ will always check
the time-stamp $T_{1}$, as  $T_{1}$ is outdated, $RC$ will considered as replay, they neglect the message request.
\subsection{Stolen Verifier Attack}

Our protocol is fully secured against stolen verifier attack. $RC$ encrypt shared key $E_{X_{c}}(X_{RS_{k}})$  using their own secret key $X_{c}$ to handle stored verifier table, so adversary does not extract anything without knowing the $X_{c}$.
\subsection{Privileged Insider Attack}

The proposed protocol successfully prevents a privilege insider attack. In the registration phase $ID_{u}$ and $A_{u} = h(N_{1}||PW_{u}||ID_{u}||CT_{u})$ are sent to $RC$, where password  $PW_{u}$ identity $ID_{u}$ a random number $N_{1}$ and cancel able template $CT_{u}$ are protected by one way hash function. So it is impossible for an insider to guess these value.
\subsection{Password Guessing Attacks}

The proposed protocol is fully secured against the password guessing attack. Suppose $RC$ take the screen shot of the user sensitive parameters like $ \{ TP_{u}, H_{u}, R, P, h(.) \aleph_{enc}(\cdot),\aleph_{dec}(\cdot)$ $ Y_{u},PID_{u},E_{u} \}$ which is stored on user smart card. Then they still requires the cancel-able transformation parameter $CT_{u}$ along with $N_1$. Moreover, an $A_{adv}$ still needs to guess identity  $ID_{u}$ and password $PW_{u}$ of user, if they unfortunately gets the $N_{1}$ and $CT_{u}$.
\subsection{Denial of Services Attack}

Our protocol is fully protected against the denial of services. $SC_{u}$ checks the validity of identity $ID_{u}$, password $PW_{u}$ and template $CT_{u}$. If $A_{adv}$ or legitimate user try to enter the incorrect values, then the $SC_{u}$ just simply cancel the request. 
\subsection{Perfect Forward Secrecy}

The proposed protocol posses the prefect forward secrecy. The shared session key $SK_{uk} = h(X^{'}_{u}||ID_{u}||SID_{k}||R_{s}||R_{u})$ incorporate a random number $R_{u}$ used by the user. Suppose if $RC$ signature $X^{'}_{c}$ is exposed to some $A_{adv}$ he will not be able to compute previously shared session keys.
\subsection{Resolve the Scalability Issues}

In previous protocol the smart card store the $AM_{uk} = (AM_{k} \oplus k^{'}_{u}) \oplus r_{u}, BM_{uk} = (BM_{k} \oplus k^{'}_{u}) \oplus r_{u}$ for every server $ 1 \leq k \leq (n + n^{'} )$, which is insufficient to store $(AM_{k})$ within smart card due to its small magnetic chip which has limited storage. In the proposed protocol there is no such parameter which stored the information of a server.
\section{Simulation tool for Formal Security Verification Using AVISPA Tool}
\label{sec: sim_avispa}

In this section, we analyze proposed protocol using formal simulation tool, for this purpose a well known and widely accepted AVISPA \cite{31armando2006avispa} tool, is used for security verification used by different authentication protocols \cite{13das2015secure, 32wazid2017design, 33srinivas2018cloud, 34chattaraj2018new}. AVISPA 
\FloatBarrier
\begin{figure}[ht]
	\centering
	\begin{minipage}[b]{0.40\textwidth}
		\includegraphics[width=1\textwidth]{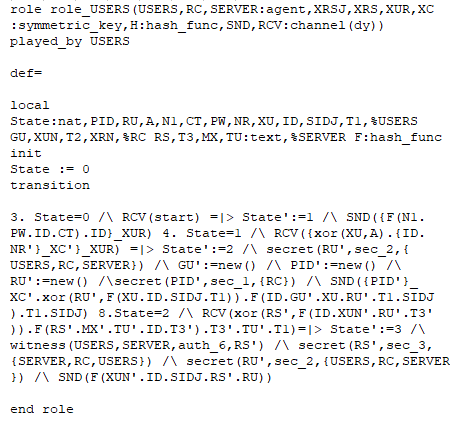}
		\caption{Role specification of user}
	\end{minipage}%
	\begin{minipage}[b]{0.40\textwidth}
		\includegraphics[width=1\textwidth]{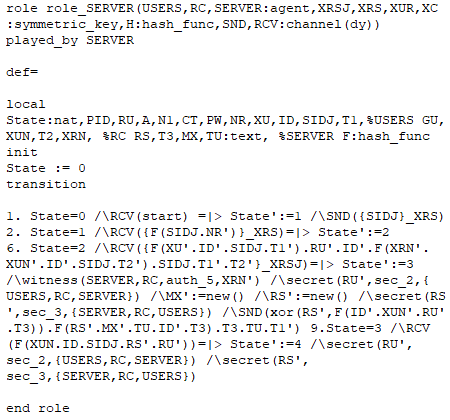}
		\caption{Role specification of server}
	\end{minipage}
\end{figure}
\FloatBarrier
implements the HLPSL language which is then translated into the intermediate formate (IF) with the help of translator known as "hlpsl2if". Four back ends are used by IF, to check security goals, is satisfied or disrupt. The output shows safe, unsafe or unsatisfactory. Details are mentioned in \cite{31armando2006avispa}. We define the three basic role i.e. role of user $U_{u}$, role of registration center $RC$ and role of server $S_{k}$ along with the session (between these participant), environment role and goals fig[5,6,7,8], are stated in HLPSL. The results of AVISPA are shown in fig[9] which tells that proposed protocol is secure against man in the middle attack as well as replay attack. The OFMC back end shows the parse time: 0.00 seconds, the search time: 42.16 seconds, the number of visited nodes is 3344 and the depth 12 plies. whereas ATSE analyzes 8 states, the translation time is 0.98 seconds. Hence, form this results it is shown our protocol provides better security against Barman et al.'s protocol \cite{21barman2018provably}. Although the search time, the translation time is slightly high compared to Barman et al.'s protocol,  because the number of visited nodes depth of proposed protocol is greater than the previous protocol.
\FloatBarrier
\begin{figure}[ht]
	\centering
	\begin{minipage}[b]{0.40\textwidth}
		\includegraphics[width=1\textwidth]{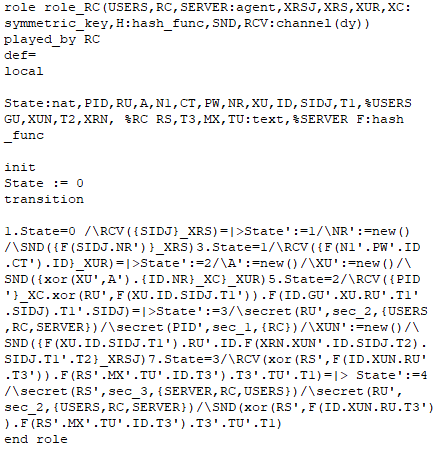}
		\caption{Role specification of Rc}
	\end{minipage}%
	\begin{minipage}[b]{0.40\textwidth}
		\includegraphics[width=1\textwidth]{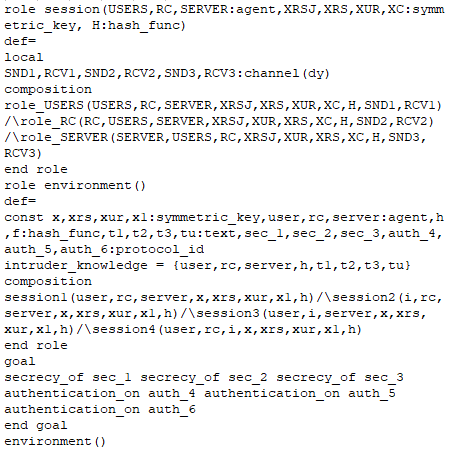}
		\caption{Role specification of Session/Goal}
	\end{minipage}
\end{figure}
\FloatBarrier
\FloatBarrier
\begin{figure}[ht]
	\centering
	\includegraphics[width=0.7\linewidth]{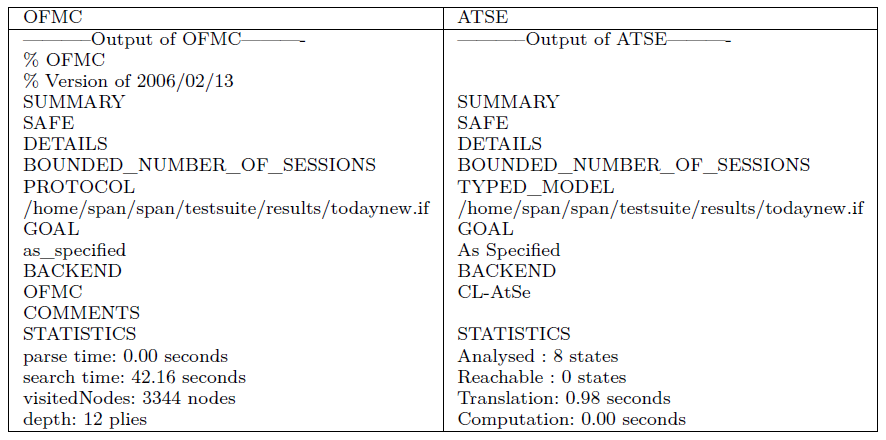}
	\caption{Results of OFMC and CL-AtSe backends}
	\label{fig:avisparesults}
\end{figure}
\FloatBarrier
\section{Performance Analysis}
\label{sec:perfor}

In this section, we evaluate the performance of the proposed protocol with other multi-server authentication protocols. The purpose of performance analysis is to check resilience of proposed protocol against various attacks.
\subsection{ Security and functionality comparisons}
The security and functionality comparison of proposed scheme with related schemes is solicited in Table \ref{TSec} under the DY and CK adversarial model as described in subsection \ref{sec:adv}. The security comparison shows that only proposed scheme provides resistance to all known attacks and fulfills related security features; whereas, all the competing schemes either lacks one or more security features or vulnerable to some security attacks.

\FloatBarrier
\begin{table*}[ht]
	\label{sec:secuirty feature}
	\centering
	 \begin{threeparttable}
		\begin{tabular}{lllllllllll}
			\hline
			Property/Feature & Our &\cite{21barman2018provably} & \cite{39chuang2014anonymous} & \cite{40amin2015novel} & \cite{41sood2011secure} & \cite{42mishra2014secure}  &  \cite{43he2014robust}  &  \cite{44lu2015robust} & \cite{45ali2017three} \\
			\hline
			$FUN_{1}$&\ding{51}  &\ding{55}  &\ding{55}  & \ding{51}  & \ding{51}  & \ding{55}  & \ding{55}  & \ding{55} & \ding{55}  \\
			$FUN_{2}$& \ding{51} & \ding{51}  & \ding{51}  & \ding{51} &\ding{55}  &\ding{51}  &\ding{51}  & \ding{51} & \ding{51} \\
			$FUN_{3}$& \ding{51} & \ding{51}  &\ding{51}  & \ding{55} & \ding{51} &\ding{51}  & \ding{51}  & \ding{51} &\ding{51} \\
			$FUN_{4}$& \ding{51} & \ding{55} & \ding{51} & \ding{51} & \ding{55} & \ding{51} & \ding{51} & \ding{51} &\ding{51}\\
			$FUN_{5}$& \ding{51} & \ding{55} & \ding{51} & \ding{51} & \ding{55} & \ding{51} &\ding{51}  & \ding{51} &\ding{51}\\
			$FUN_{6}$& \ding{51} & \ding{51} & \ding{51} & \ding{51} & \ding{55} & \ding{51} & \ding{51} & \ding{51} &\ding{51}\\
			$FUN_{7}$& \ding{51} & \ding{55} & \ding{51} & \ding{51} & \ding{51} & \ding{55} & \ding{51} & \ding{51} &\ding{51}\\
			$FUN_{8}$& \ding{51} &  \ding{51}&  \ding{51}& \ding{51} & \ding{51} & \ding{51} & \ding{51} & \ding{51} &\ding{51}\\
			$FUN_{9}$& \ding{51} & \ding{51} & \ding{51} & \ding{55} & \ding{55} & \ding{51} & \ding{51} & \ding{51} &\ding{51}\\
			$FUN_{10}$& \ding{51} & \ding{51} & \ding{51} & \ding{51} & \ding{51} & \ding{51} & \ding{55} & \ding{55} &\ding{51}\\
			$FUN_{11}$&\ding{51}  &  \ding{51}& \ding{51} & \ding{55} & \ding{55} & \ding{51} & \ding{51} &\ding{55}& \ding{55} \\
			\hline
\end{tabular} \label{TSec}
	
	\begin{tablenotes}
	\item \tiny $FUN_{1}$: user anonymity violation and untraceability; $FUN_{2}$: three-factor security feature; $FUN_{3}$: error detection mechanism; $FUN_{4}$:  participant having mutual authentication; $FUN_{5}$: exchange of session key; $FUN_{6}$: password update security; $FUN_{7}$: resistance against stolen smart card attack; $FUN_{8}$: resistance against offline password guessing; $FUN_{9}$: resistance against replay attack; $FUN_{10}$:resistance against forgery attack; $FUN_{11}$: resistance against privileged-insider attack. 
	\item \ding{51}: a protocol safeguard the security functionality feature; \ding{55}: a protocol is lack of the security functionality feature.
		\end{tablenotes}
		 \end{threeparttable}
	\caption{Security and functionality features comparison}

\end{table*}

\subsection{Computation cost}

In this subsection, we compare our protocol with the existing multi-server authentication protocols considering the computation cost of login and authentication phases. The following notation used for computation cost describe below:
\begin{description}
	\item[$\bullet$] $F_{H_{cost}}$: one-way cryptographic hash cost  
	\item[$\bullet$] $F_{BH_{cost}}$: bio-hashing cost
	\item[$\bullet$] $F_{FE_{cost}}$: fuzzy extractor cost  
	\item[$\bullet$] $F_{FCS_{cost}}$: fuzzy commitment cost
	\item[$\bullet$] $F_{ECM_{cost}}$: ecc point multiplication cost
	\item[$\bullet$] $F_{ASYM_{cost}}$: asymmetric key encryption/decryption cost
	\item[$\bullet$] $F_{Enc_{cost}}$:  cost of block cipher encryption
\end{description}
The experimental results disclosed in \cite{18reddy2017design}, we choose $F_{H_{cost}}$ = 0.0023 ms, $F_{Enc_{cost}}$ = 0.0046 ms, $F_{ECM_{cost}}$ = 2.226 ms and $F_{ASYM_{cost}}$ = 0.0046 ms. Furthermore, $F_{FE_{cost}} = F_{ECM_{cost}}$, we also assume $F_{BH_{cost}} = F_{ECM_{cost}}$ and $F_{FCS_{cost}} = F_{ECM_{cost}}$. Although our protocol has slightly high computation cost compared to Barman et al \cite{21barman2018provably}, but the security level of our protocol is high. The comparison are briefly shown in Table \ref{TComp}.
\begin{table*}[ht]
	\label{sec:compCost}
	\centering
	\scalebox{0.9}{
		\begin{tabular}{lllllllll}
			\hline
			Protocol &  Login phase & Authentication phase & Total cost &  Rough estimation (ms) \\
			\hline 
			Chuang-Chen\cite{39chuang2014anonymous} & $4C_{h}$ & $13C_{h}$ & $17C_{h}$  & 0.0391   \\
			Amin-Biswas\cite{40amin2015novel} & $C_{bh} + 4C_{h}$  &  $14C_{h}$  & $ C_{bh} + 18C_{h} $  & 2.2674   \\
			Sood\cite{41sood2011secure} & $7C_{h}c$ & $24C_{h}$  & $31C_{h}$  & 0.0713   \\
			Mishra\cite{42mishra2014secure} & $6C_{h} $ & $12C_{h}$  & $18C_{h} $ &  0.0414   \\
			He-Wang\cite{43he2014robust}  & $3C_{h} + 2C_{ecm}$  &  $18C_{h} + 6C_{ecm}$ & $21C_{h} + 8C_{ecm}$ &  17.856  \\
			Lu\cite{44lu2015robust}& $C_{bh} + 4C_{h}$ & $11C_{h}$ &   $C_{bh} + 15C_{h}$ &  2.2605  \\
			Ali-Pal\cite{45ali2017three} & $6C_{h} + C_{asym} + C_{bh}$ &  $7C_{h} + C_{asym}$ & $ 13C_{h} + C_{bh} + 2C_{asym} $ &  2.2651   \\
			Barman\cite{21barman2018provably} & $C_{fcs} + 6C_{h}$ & $ 11C_{h}$  & $C_{fcs} + 17C_{h}$ &  2.2651   \\
			Our & $C_{fcs} + 6C_{h}$ & $ 13C_{h}+1C_{{E_{X_{c}}}} +1C_{{h}_{D_{X_{c}}}}$  & $C_{fcs} + 19C_{h} +1C_{{E_{X_{c}}}} +1C_{{h}_{D_{X_{c}}}}$ &  2.2789\\
			\hline
		\end{tabular}
	} 
	\caption{Computation costs comparison} \label{TComp}
\end{table*}

\subsection{Communication cost }

In this subsection, we evaluate and compare the communication cost of proposed with existing protocols. During the login and authentication phases, the communication cost is computed by the total number of bits which is transmitted  to other parties in the network, over a protected channel. We are assuming the "SHA-1" hash function which has the cost of 160 bits\cite{35burrows2015secure}, in the symmetric key encryption/decryption, has the cost of 256 bits of length\cite{36kumar2019sebap}, time stamp is 32 bits of length, an elliptic curve point $P = (P_{a},P_{b})$ is 160 length of bits, where $P_{a}$ and $P_{b}$ is x and y coordinate of P point. Furthermore the security of RSA\cite{37rivest1978method} public key cryptosystem is 1024-bit which is comparable to ECC (elliptic curve cryptography) of 160-bits of length \cite{38barker2012recommendation}. In the proposed protocol, the communication cost for the login request message $\{DID_{u}, H_{u}, G_{u}, T_{1}, SID_{k}\}$, which is transmitted from a user $U_{u}$ to the$RC$ has cost of (160+160+160+32+32) = 544 bits of length
and the message $\{E_{X_{RS_{k}}}(X^{'}_{u}, R_{u}, ID_{u}, H_{R_c}, SID_{k}, T_{1}), SID_{k}, T_{2}\}$ transmitted to server $S_{k}$ from $RC$ is (256+32+32) = 332 bits and the message  transmitted to $RC$ from server $S_{k}$ is $\{M_{x}, H^{''}_{R_{c}}, T_{3}, T_{u},\}$ (160+160+32+32) = 384 bits and message  transmitted to $U_{u}$ from $RC$ is $\{M_{x}, H^{''}_{R_{c}},T_{3},T_{u}, RID_{u }\}$ (160+160+32+32+160) = 544 bits hence, the total number of bits for communication is (544+332+384+544) = 1804 bits. The comparison results are  shown in Table \ref{TComm}.
\begin{table*}[ht]
	\label{sec:comCost}
	\scalebox{0.9}{
		\begin{tabular}{lllllllll}
			\hline
			protocol & Cost in login phase & Cost in authentication phase &  Total cost&  Communication mode \\
			\hline 
			Chuang-Chen\cite{39chuang2014anonymous} &512   &512   & 1024   &  $ U_{u} \xrightarrow{} S_{k}, S_{k} \xrightarrow{} U_{u}$  \\
			Amin-Biswas\cite{40amin2015novel} & 768   &1152  & 1920   & $ U_{u} \xrightarrow{} {}MS, MS \xrightarrow{} PS, PS \xrightarrow{} U_{u}$   \\
			Sood\cite{41sood2011secure} & 896   &1216  & 2112  &  $ U_{u} \xrightarrow{} S_{k}, S_{k} \xrightarrow{} CS, CS \xrightarrow{} S_{k}, S_{k} \xrightarrow{} U_{u}, U_{u} \xrightarrow{} S_{k}$  \\
			Mishra\cite{42mishra2014secure} & 640 & 640  & 1280 &  $U_{u} \xrightarrow{} S_{k}, S_{k} \xrightarrow{} U_{u}, U_{u} \xrightarrow{} S_{k}$  \\
			He-Wang\cite{43he2014robust} & 640  &2880  & 3520  &  $ U_{u} \xrightarrow{} S_{k}, S_{k} \xrightarrow{} RC, RC \xrightarrow{} S_{k}, S_{k} \xrightarrow{} U_{u}, U_{u} \xrightarrow{} S_{k} $ \\
			Lu\cite{44lu2015robust} & 672   & 554  & 1226  &   $U_{u} \xrightarrow{} S_{k}, S_{k} \xrightarrow{} U_{u}, U_{u} \xrightarrow{} S_{k} $  \\
			Ali-Pal\cite{45ali2017three} & 1344  & 320  & 1664  &   $U_{u} \xrightarrow{} S_{k}, S_{k} \xrightarrow{} U_{u} $ \\
			Barman\cite{21barman2018provably} & 544 & 1164 & 896 & $U_{u} \xrightarrow{} S_{k}, S_{k} \xrightarrow{} U_{k} $  \\
			Our & 544 & 1260 & 1804 & $U_{u} \xrightarrow{} RC, RC \xrightarrow{} S_{k}, RC \xleftarrow{} S_{k}, U_{u} \xleftarrow{} RC $ \\
			\hline
		\end{tabular}
	}
	\caption{Communication cost comparison} \label{TComm}
\end{table*}
\FloatBarrier
\section{Conclusion}
\label{sec:con}
Very recently Barman et al. presented a provably secure multi-server authentication protocol using fuzzy commitment. The authors claimed that their protocol provides various security services like privacy preservation of user’s identity and biometric data, mutual authentication and session key establishment between user and servers. Furthermore, barman et al. also claimed that their protocol is secure against all known attacks.  However, the analysis in this paper shows that Barman et al.'s protocol cannot withstand user anonymity violation as well as impersonation attack alongwith the scalability issues. Then we proposed an improved and enhanced protocol to fix the weaknesses of Barman et al.'s protocol. The proposed protocol is more robust than Barman et al. and related protocols which is evident from rigorous formal and informal security analysis. We have also validated the security of the proposed protocol by simulation in popular and widely accepted security analysis tool AVISPA.

\bibliographystyle{spmpsci}

\end{document}